\documentclass[aps, prd, amsmath,amssymb,superscriptaddress,preprintnumbers,twocolumn,
 amsmath,amssymb,
 aps,
]{revtex4-2}

\usepackage{graphicx}
\usepackage{dcolumn}
\usepackage{bm}
\usepackage{cancel}

\usepackage[usenames, dvipsnames]{xcolor}
\usepackage{tikz}
\usetikzlibrary{shapes}
\usepackage[co mpat=1.1.0]{tikz-feynman}
\usepackage{float}

\usepackage{hyperref}

\usepackage{youngtab}
\usepackage{ytableau}

\usepackage[T1]{fontenc}
\usepackage{amsfonts}
\usepackage{amssymb}
\usepackage[utf8]{inputenc} 
\usepackage{graphicx} 
\usepackage{braket}
 \usepackage{fancyhdr}
\usepackage{mathrsfs}
\usepackage{setspace}
\usepackage{amsfonts}
\usepackage{amssymb}
\usepackage{amsmath}
\usepackage{bbm}
\usepackage{epsfig}
\usepackage{latexsym}
\usepackage{color}
\usepackage{nicefrac}
 \usepackage{slashed}
 \usepackage{multirow}
 \usepackage{comment}
 \usepackage{hyperref}
\usepackage{slashed}
\usepackage{array}
\usepackage{booktabs}

\def\be{\begin{equation}}
\def\ee{\end{equation}}

 \newcommand   \cO {\mathcal{O}}
\def\sn{{\rm sn}}
\def\cn{{\rm cn}}

\def \ELE {{\mathbb{E}}}
\def \KK {{\mathbb{K}}}
\newcommand{\ba}{\begin{eqnarray}}
\newcommand{\ea}{\end{eqnarray}}

\def \T {{\cal T}}

\newcommand{\order}[1]{ \mathcal{O}\left({#1}\right) }

\begin{document}


\title{Semiclassical spectrum of the Ising CFT}


\author{Oleg Antipin}%
\email{oantipin@irb.hr}
\affiliation{ Rudjer Boskovic Institute,
  Division of Theoretical Physics,
  Bijeni\v cka 54, 10000 Zagreb, Croatia}
  
\author{Jahmall Bersini}
\email{jahmall.bersini@unibe.ch}
\affiliation{\mbox{Albert Einstein Center for Fundamental Physics,
Bern University,
Sidlerstrasse 5, CH-3012 Bern, Switzerland}}

\author{Jacob Hafjall}%
\email{jahaf21@student.sdu.dk}
\affiliation{\mbox{Quantum Theory Center ($\hslash$QTC) at IMADA \& D-IAS, Southern Denmark Univ., Campusvej 55, 5230 Odense M, Denmark}}

\author{Giulia Muco}%
\email{giulia@qtc.sdu.dk}
\affiliation{\mbox{Quantum Theory Center ($\hslash$QTC) at IMADA \& D-IAS, Southern Denmark Univ., Campusvej 55, 5230 Odense M, Denmark}}

\author{Francesco Sannino}
\email{sannino@qtc.sdu.dk}
\affiliation{\mbox{Quantum Theory Center ($\hslash$QTC) at IMADA \& D-IAS, Southern Denmark Univ., Campusvej 55, 5230 Odense M, Denmark}}
\affiliation{Dept. of Physics E. Pancini, Universit\`a di Napoli Federico II, via Cintia, 80126 Napoli, Italy}

 \begin{abstract}
We develop a semiclassical framework to determine scaling dimensions of neutral composite operators in scalar conformal field theories. For the critical Ising $\lambda\phi^4$ theory in $d=4-\epsilon$, we obtain the full spectrum of composite operators built out of $n$ fields transforming according to the various Lorentz representations to next-to-leading order in the double--scaling limit $n\rightarrow \infty$ and $\lambda \rightarrow 0$ with $\lambda n$ fixed. At any given order, the semiclassical expansion resums an infinite number of Feynman diagrams. Combining our results with existing perturbative computations further yields the complete five-loop scaling dimensions in the $\epsilon$-expansion for the family of $\phi^n$ operators. Finally, extrapolating the semiclassical results to three dimensions leads to a competitive alternative to existing methodologies. 
 \end{abstract}

 \maketitle
 \section{Introduction}
Conformal field theories (CFT)s play a major role in describing critical phenomena in nature, from condensed matter to fundamental interactions. Scaling dimensions are relevant CFT data associated with the two-point function of the primary operators of the theory. In the past decades a number of methodologies have been developed to tackle the important dynamics related to composite operators. These range from the use of larger symmetries, such as supersymmetry, to large quantum numbers  \cite{Hellerman:2015nra, Komargodski:2012ek} expansions, bootstrap \cite{Rattazzi:2008pe}, as well as numerical approaches. These approaches, either perturbative or nonperturbative, fall short of determining the conformal dimensions of neutral composite operators built out of a large number $n$ of fields. 
To better appreciate the challenge, even in a weakly-coupled field theory with a single coupling $\lambda \ll 1$, the perturbative expansion for the scaling dimension of these operators is controlled by $\lambda n$ and therefore it breaks down for $\lambda n$ of order unity \cite{Badel:2019oxl}. Moreover, operator mixing under renormalization group (RG) flow leads to an anomalous dimension matrix whose size grows fast with $n$, making it hard to determine the physical spectrum (scaling dimensions) of primary operators. Beyond CFTs, determining the RG functions associated with higher-dimension operators is an important problem in the renormalization of fundamental and effective theories. This is, for example, particularly relevant for the Standard Model effective field theory \cite{Buchmuller:1985jz, Grzadkowski:2010es, Brivio:2017vri}, where higher-dimension operators are included to parametrize new physics effects at colliders.
Perturbation theory also fails when studying multi-boson production processes, e.g., $1 \to n$, $2 \to n$, in generic weakly-coupled quantum field theories (QFT)s \cite{Goldberg:1990qk, Cornwall:1990hh, Son:1995wz}. This problem also afflicts Standard Model processes involving a large number of W, Z, and Higgs bosons \cite{Khoze:2014zha}. These issues can be traced back to the actual computation of multi-legged amplitudes, where the large number of loops and legs leads to an exponential proliferation of Feynman diagrams, severely hampering the efforts in solving for the dynamics of QFTs. 

The overarching goal of this work is twofold. First, to introduce a semiclassical methodology apt at determining the scaling dimensions of composite operators beyond perturbation theory and without the use of Feynman diagrams, consolidating and going beyond the work in \cite{Antipin:2024ekk}. Specifically, while \cite{Antipin:2024ekk} calculated the leading semiclassical order, we here determine the leading quantum correction. 
Second, we will determine, to next-to-leading order in the semiclassical expansion, the full spectrum of operators in the critical (Ising) $\lambda \phi^4$ theory in $d=4-\epsilon$ dimensions
\begin{equation}
    \mathcal{L} =\frac{1}{2}(\partial \phi)^2 - \frac{\lambda}{4} \phi^4 \,,
\end{equation}
where the Wilson-Fisher fixed point coupling occurs at
\be \label{chiodofisso}
\lambda_* =\frac{8 \pi ^2 \epsilon }{9} +  \frac{136 \pi ^2 \epsilon ^2}{243} + \order{\epsilon^3} \ .
\ee
As we shall see, when extrapolated to $\epsilon =1$, our results exhibit reasonable agreement with the conformal bootstrap for the low-lying spectrum of the three-dimensional Ising CFT while providing for the first time a controlled framework for the higher-dimensional operators. Remarkably, while at the leading semiclassical order, the full spectrum of the theory is degenerate \cite{Antipin:2024ekk}, the inclusion of the next-to-leading order (NLO) lifts this degeneracy. This allows us to distinguish operators belonging to different Lorentz representations, unveiling the richness of the spectrum and its previously observed surprising degeneracies \cite{Kehrein:1992fn, Kehrein:1994ff}. At the same time, the NLO calculation unveils the limitations of the approach as the relevant classical trajectory develops instabilities in the strongly coupled regime.

In the free field theory limit, the composite operators can be generically denoted as $\partial^s \Box^p \phi^n$. With this notation, we intend the appropriate linear combination of terms with $n$ fields, $2p$ fully contracted and $s$ uncontracted derivatives, such that the resulting operator is a conformal primary. There can be different primaries for the same values of $s$, $p$, and $n$. Operators transforming according to nonfully symmetric representations of the Lorentz group will be further specified via a subscript in the partial derivative symbol. In the interacting theory, these operators undergo mixing under renormalization, and the primary operators become certain linear combinations of free field theory primaries that diagonalize the anomalous dimension matrix. For simplicity, we here schematically denote the interacting primaries by those which they reduce to in the corresponding $\epsilon \to 0$ free field theory limit. Via the state-operators correspondence, conformal dimensions can be mapped into the energy spectrum of the corresponding states on the cylinder $\mathbb{R}\times S^{d-1}$. Building on the initial intuition put forward in \cite{Cuomo:2024fuy, Antipin:2024ekk}, we here demonstrate that the full spectrum of scalar composite operators maps into the energy spectrum $E_{n,\{q_\ell\}}$ of the states obtained by quantizing the periodic homogeneous solutions of the classical equation of motions on the cylinder. Their energy is computed via a semiclassical expansion in the limit $n\rightarrow \infty$, $\lambda \rightarrow 0$, and $ \lambda n $ fixed, leading to the desired scaling dimensions 
\begin{equation} \label{dsl1}
    \Delta_{n,\{q_\ell\}} = r E_{n,\{q_\ell\}} = n \sum_{i=0} \frac{{C}_i(\lambda n,\{q_\ell\})}{n^i} \ ,
\end{equation} 
where $r$ is the radius of $S^{d-1}$ and $\{q_\ell\}$ is an infinite set of integers specifying the states and the corresponding composite operators $\partial^s \Box^p \phi^n$. The leading coefficient $C_0$ has been computed in \cite{Antipin:2024ekk} and reads 
\begin{equation}
\label{C0}
  n C_{0}=  \frac{2\pi^2  m \,(1-m) } {\lambda \,(1-2 m)^2} \ .
\end{equation}
where $m$ is related to $\lambda n$ by inverting the Bohr-Sommerfeld quantization condition $I = 2\pi n$ where the action variable $I$ reads
\begin{align}
I = \frac{16 \pi^2  }{3  \lambda (1-2 m)^{3/2}} \left[ (2m-1)\ELE(m) + (1-m)\KK(m) \right]\,.\nonumber
\end{align}
with $\KK(m)$ and $\ELE(m)$ the complete elliptic integrals of the first and second kind. The NLO term can be determined using the framework developed in \cite{Gutzwiller:1971fy,dhn,Dashen:1974ci, Beccaria:2010zn} for the quantization of periodic solutions in QFT. Following the derivation in App.~\ref{app1}, which uses Refs.~\cite{Sanchez, floquet, Antipin:2025rsr, book1, Pawellek:2008st, arscott},  we obtain

\begin{align} \label{labello}
       &  C_1 = \frac{5 \pi\sqrt{1-2m}  -24\mathbb{E}(m)}{4(1-2m) \mathbb{K}(m)} + \frac{28-67m+19m^2 }{8(1-2m)^2}  \nonumber \\ &+ \frac{1}{8 \sqrt{1-2m }\KK(m)} \left(\sum_{\ell=2}^\infty \sigma(\ell) +2\sum_{\ell=1}^\infty q_\ell \nu_\ell \right)\,, 
    \end{align}
where
{\small{
    \begin{align}\label{C1} 
         \sigma(\ell)&=   (\ell+1)^2 \nu_{\ell}  +\frac{2}{\ell(1-2 m)^{3/2}} \Bigg(6 \left(\ell^2+\ell+1\right) (2 m-1) \ELE(m) \nonumber \\ &+ \left( 6+4 \ell  -2 (\ell+3) \ell^3+(2 \ell (\ell+1) (2 \ell-1) (2 \ell+5)-15) m\right. \nonumber \\  & \left.+  (9-4 \ell (\ell+1) (2 \ell (\ell+2)-1)) m^2\right)\KK(m) \Bigg)  \ .
    \end{align}}}
Here, $\nu_{\ell}$ are the \emph{stability angles} of the fluctuations around the classical solution and read
\be \label{stab2}
 \nu_{\ell} = 2\pi - 4 i \KK(m)  ( Z(\alpha_+\,|\,m)+\, Z(\alpha_-\,|\,m)) \ ,
\ee
where $Z$ denotes the Jacobi Zeta function and $\alpha_\pm$ solve 
\begin{align}\label{eq:Translos}
 \sn^2(\alpha_\pm |\,  m)&= \pm\frac{1}{2} \sqrt{1-\frac{1}{3} \ell (\ell+2) (\ell (\ell+2)-2) (1-2 m)^2} \nonumber \\ & +\frac{1}{2}+\frac{1}{6} \ell (\ell+2) (2 m-1)\,.
\end{align}
with $\sn(z|m)$ the Jacobi elliptic sine function.
The above constitute the main technical result of this letter. The sums over $\ell$ in $C_1$ are convergent and can be evaluated numerically for any value of $m=m(\lambda n)$ or analytically in the small/large $\lambda n$ regime. The leading semiclassical order ${C}_0$ is $q_\ell$ independent. Remarkably, this implies that the full spectrum of composite operators of the theory is degenerate at this order. We now proceed by analyzing the implications of our main result.
\section{The small $\lambda n$ expansion} 
The general (asymptotic) perturbative expansion of $\Delta_{n,\{q_\ell\}} $ assumes the form
\begin{equation} \label{roperto}
  \Delta_{n,\{q_\ell\}} \sim \sum_{k=0}^\infty P_k(n)\epsilon^k   \,, \ \ \ \ \ P_k(n) =\sum_{i=0}^{k+1} c_{ik} \ n^{k+1-i} \,,
\end{equation}
where the $k$-loop coefficient $P_k(n)$ is a polynomial of degree $k+1$ in $n$. By comparing the above to Eq.\eqref{dsl1}, we deduce that the perturbative expansion of the $C_i$ coefficients reads
\be
C_i \sim \sum_{k=0}^\infty c_{ik} \ (\epsilon n)^{k} \ ,
\ee
and, therefore, the coefficients $c_{ik}$ can be read off from the small $\epsilon n$ (i.e. small $\lambda n$) expansion of $C_i$. In other words, $C_0$ resums the terms with the leading power of $n$ at any loop order, $C_1$ resums those with the next-to-leading power, and so on. By expanding our results in Eq. \eqref{C0} and Eq. \eqref{C1} around $\epsilon n = 0$, we obtain  
\onecolumngrid
    \begin{align} \label{mastero}
\Delta_{n,\{q_\ell\} } &= \textcolor{red}{n} + \textcolor{blue}{\sum_{\ell=1}^{\infty} q_\ell \ell} + \frac{1}{6}   \left[\textcolor{red}{n^2}-\textcolor{blue}{2\left(2+\sum_{\ell=1}^{\infty} \frac{(\ell-1)q_\ell}{\ell+1}\right)}\textcolor{blue}{n}+  {\order{n^0}}\right] \epsilon \nonumber \\ &- \frac{1}{324}  \left[\textcolor{red}{17 n^3}  - \textcolor{blue}{\left(67 +3 \sum_{\ell=1}^{\infty} \frac{ (\ell-1) \left(17 \ell^4+78 \ell^3+135 \ell^2+98 \ell+12\right) q_\ell}{\ell (\ell+1)^3 (\ell+2)}\right)}\textcolor{blue}{n^2} + \order{ {n}, {n^0}}\right]\epsilon ^2  + \order{\epsilon^3} \ ,
\end{align}
\twocolumngrid
where we color-coded in red and blue the $C_0$ and $C_1$ contributions.  
Explicit $7$-loop results (i.e., the values of the coefficients $c_{0k}$ and $c_{1k}$ for $k=1,\dots,7$) for any finite set of integers $\{ q_\ell\}$ are available in the ancillary Mathematica file \cite{ancillary}. 
The family of primaries corresponding to the ground state (i.e. $q_\ell = 0 \ \forall \ \ell$) is typically denoted schematically as $\phi^n$ and includes all the lowest Lorentz scalars of the theory. For arbitrary $n$, their scaling dimensions, henceforth denoted as $\Delta_n$, have been determined in the $\epsilon$-expansion up to three-loop order in \cite{Antipin:2024ekk}, and one can easily check that these results match Eq. \eqref{mastero} for $q_\ell = 0$. A relevant application of our findings is the possibility of deriving new perturbative results by combining semiclassical and conventional diagrammatic computations. Specifically, at the $k$-loop order, our NLO calculation determines two of the $c_{ik}$ coefficients of the polynomial $P_k(n)$ appearing in Eq.~\eqref{roperto}, and the remaining $k$ coefficients can be fixed by matching onto existing $k$-loop perturbative data for $k$ distinct values of $n$. In particular, by employing the most up-to-date diagrammatic results~\cite{Henriksson:2025hwi, Schnetz:2016fhy, Schnetz:2022nsc, Bednyakov:2021ojn, Kompaniets:2017yct}, we obtain the complete $\mathcal{O}(\epsilon^5)$ (five-loop) expression for the scaling dimension of $\phi^n$ which for all $n \ge 8$ improves upon previously available lower-order computations
{\small\begin{align}
 \label{fiveloops}
&\Delta_n =\left(1-\frac{\epsilon}{2}\right)n +\frac{n}{6} (n-1)  \epsilon -\frac{n}{324} \left(17 n^2-67 n+47 \right) \epsilon ^2 \nonumber \\ & + (0.0321502 n^4-0.123379 n^3+0.0734761 n^2+0.0270975 n)\epsilon^3  \nonumber \\ &+(-0.0254558 n^5+0.113814 n^4-0.201973 n^3+0.348649 n^2 \nonumber \\ &-0.239199 n)\epsilon^4 + (0.0231664 n^6-0.121078 n^5+0.327421 n^4\nonumber \\ &-0.53835 n^3-0.0673349 n^2+0.389004 n) \epsilon^5 \,.
\end{align}}
\noindent The analytic form can be found in the ancillary file.
The stability angles are ordered as $\nu_{\ell} < \nu_{\ell+1}$, and therefore the first excited state is obtained by exciting a single $\ell=1$ mode (i.e., setting $q_\ell = \delta_{1\ell}$, with $\delta_{1\ell}$ the Kronecker delta). Since $\nu_1=\T$, the corresponding conformal dimension increases by one unit, yielding a descendant state corresponding to acting with the conformal raising operator $\partial_\mu$. Therefore, a necessary criterion for an excited state to be a primary is $q_1 =0$. Moreover, given any primary operator, the corresponding full conformal multiplet is generated by exciting the $\ell=1$ mode repeatedly. Next, we observe that exciting a single mode with $\ell=s>1$, one obtains families of spin $s$ operators, schematically of the form $\partial^s\phi^n$. Their dimension is obtained by setting $q_\ell = \delta_{s \ell}$ in Eq. \eqref{mastero} and we checked that the results match known $1$-loop calculations, which are available for $s=2,3,4,5$ \cite{Kehrein:1994ff} (see also \cite{Henriksson:2022rnm} for a comprehensive review of the Ising CFT data). These operators are generically primaries, but for specific values of $n$ and $s$ (e.g., $n=4$, $s=5$), they may be descendants.
By continuing this construction, one can see that by exciting multiple modes, we can reproduce the full spectrum of primary operators of the Ising CFT. In essence, exciting an $\ell$ mode amounts to adding $\ell$ derivatives in the $(\ell/2,\ell/2)$ representation of $SO(4)$ and, to identify the resulting operators, one has to consider the corresponding decomposition into irreducible Lorentz representations. For instance, by exciting two $\ell=2$ modes we have to decompose the product of two $\left(1,1\right)$ representations using
\be
1 \otimes 1 = 2 \oplus 1 \oplus 0 \ ,
\ee
Accordingly, the Ising CFT contains the four operators  $\partial^4 \phi^n$, $\partial^2 \Box \phi^n$, $\Box^2 \phi^n$, and $\partial_{\scalebox{0.2}{\footnotesize \yng(2,2)}}^4 \phi^n$, which, up to the computed semiclassical order, have the same scaling dimensions obtained by setting $q_\ell = 2 \delta_{2\ell}$ in Eq. \eqref{mastero}. This result was again checked by comparing it to the $1$-loop results in \cite{Kehrein:1992fn, Kehrein:1994ff}. Intuitively, the spin zero operator is built by contracting four derivatives (two for each $\ell=2$ mode), and so will schematically be of the form $\Box^2 \phi^n$, i.e., $s=0, \ p=2$. Analogously, the spin two and four operators have $s=2, \ p=1$ and $s=4, \ p=0$, respectively.  Finally, we exhaust the list of available perturbative results by further comparing to \cite{Kehrein:1994ff} the dimension of the primaries obtained by exciting  $\ell=2$ and $\ell=3$ mode ($q_\ell = \delta_{2\ell}+ \delta_{3\ell}$), which, since $3/2 \otimes 1 = 5/2 \oplus 3/2 \oplus 1/2$, have $s=\{5,3,1\}$ uncontracted derivatives. 

We remark that when rewriting $\epsilon$ in terms of the renormalized coupling, the results above for the scaling dimensions extend away from the conformal regime, where they become anomalous dimensions in dimensional regularization. 

\section{On the large $\lambda n$ expansion} 
By expanding our results in Eq. \eqref{C0} and Eq. \eqref{C1} for large $\lambda n$ limit and focusing on the ground state ($q_\ell=0 \ \forall \ \ell$), we  obtain 
    \begin{align}
\Delta_n &=  \left(\frac{ \sqrt{\frac{3 \pi }{2}} \Gamma \left(\frac{3}{4}\right)}{\Gamma \left(\frac{1}{4}\right)}n \epsilon\right)^{\frac{4-\epsilon }{3-\epsilon }}\left(\frac1\epsilon+ \alpha + \order{\epsilon}\right) + \order{n^{\frac{2-\epsilon }{3-\epsilon }}} \ ,
\end{align}
where $\alpha$ is an integral over a function stemming from the expansion of the stability angles in the large $\lambda n$ limit.
Interestingly, the NLO correction modifies the large $n$ behavior from $\Delta_n \sim n^{4/3}$ found in \cite{Antipin:2024ekk} to
\be
\lim_{n \to \infty} \Delta_n \sim n^{\frac{d}{d-1}} \ .
\ee
Intriguingly, this result mimics the generic nonperturbative behavior of the scaling dimension of the lowest operator with charge $n$ in generic CFTs with global symmetries \cite{Hellerman:2015nra}. However, in our case, there is more to it. In fact, the integrand defining $\alpha$ has a branch cut, which makes  $\alpha$ multivalued. Its presence can be understood by noting that $\nu_\ell$ in \eqref{stab2} becomes complex for any $\ell$ in the range
\be \label{ranger}
 \sqrt{\frac{4-5 m}{1-2 m}}\le \ell + 1\le \sqrt{2}\sqrt{1+\frac{ \sqrt{1-m+m^2}}{1-2 m}} \ .
\ee
Whenever an integer value $\ell$ falls in the above range, the corresponding stability angles become complex, signaling an instability of the associated classical orbit. In such a case, the computation breaks down, and one may need to take into account additional (generically inhomogeneous) saddle points. Physically, when $\ell$ satisfies the above, the spectral parameter of the finite-gap fluctuation operator around the classical solution takes values outside the allowed bands of the spectrum. For small values of $m$ (i.e., small values of $\lambda n$), no integer $\ell$ satisfies the above condition. However, as $\lambda n$ increases, the $\ell=2$ mode eventually enters the unstable orbit region for $m=3/8$, corresponding to $\lambda n  \approx 50 $. In general, for $3/8<m<\frac{1}{65} \left(4 \sqrt{61}+1\right)$, there are ranges of values of $m$ such that there is a single complex stability angle for a certain integer $\ell$. For $m \ge \frac{1}{65} \left(4 \sqrt{61}+1\right)$ ($\lambda n \gtrapprox 10856$), there is always at least one complex stability angle. Finally, in $m \to 1/2$ ($\lambda n \to \infty$) limit, where the stability angles tend to a continuum, the complex ones fill the whole region $\sqrt{3/2}< \ell \sqrt{1-2m} < 3^{1/4}$.
Therefore, for certain values of $\lambda n$ in the range $50 \lessapprox \lambda n \lessapprox 10856$ and for all $  \lambda n \gtrapprox 10856 $, one has to investigate further the saddle point analysis. Since a precise analysis of the instability would require performing an involved Lefschetz thimble analysis of the path integral \cite{Dunne:2015eaa}, we leave it to future investigations.  In particular, it would be interesting to understand the relation between the instability as a Stokes phenomenon, operator mixing, and chaotic properties of the CFT spectrum.

\section{Ising CFT in three dimensions} 
\begin{figure}[t!]
\centering
\includegraphics[width=0.51\textwidth]{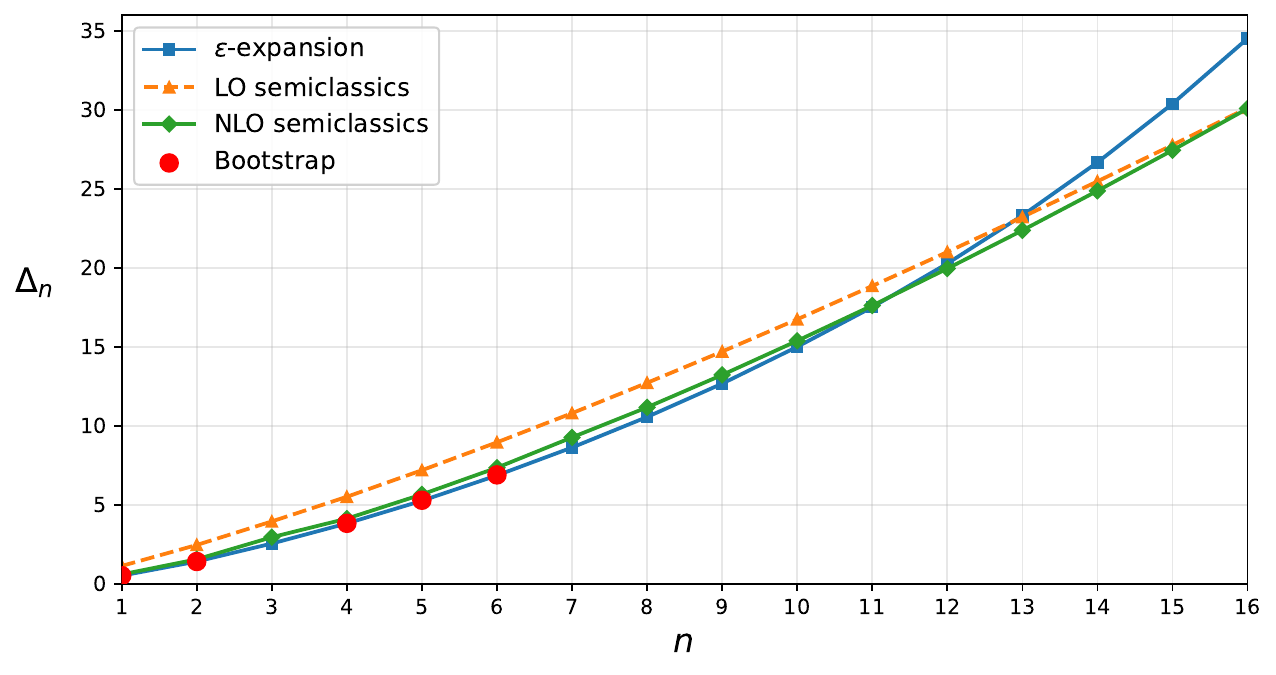} 
	\caption{$\Delta_n$ for $n \le 16$ obtained via numerical bootstrap,  $\epsilon$-expansion,  and semiclassics at LO and NLO.} 
	\label{Moneyplot}
\end{figure}

As a relevant application of our methodology, we now present the conformal dimensions $\Delta_n$ associated with the   $\phi^n$ operators for the physical Ising CFT in $d=3$.  In Fig.~\ref{Moneyplot} we compare our NLO semiclassical predictions to the results obtained via the conformal bootstrap and the $\epsilon$-expansion.  

For $n=1,2,4$, $\Delta_n$ has been measured experimentally but with large uncertainties \cite{Henriksson:2022rnm}. The theoretical estimates obtained via Monte Carlo simulations and conformal bootstrap are more accurate and fall within the experimental bounds. More generally, bootstrap results are available for $n=1,2,4,5,6$ \cite{Simmons-Duffin:2016wlq, Henriksson:2022gpa}. There is no primary operator with $n=3$. The $\epsilon$-expansion results have been obtained as follows: for $n=1$ we considered the conformal-Borel resummed $6$-loop result of \cite{Kompaniets:2017yct}, for $n=2,4$, the hypergeometric–Meijer resummation of the $7$-loop results of \cite{Schnetz:2016fhy} performed in \cite{Shalaby:2020xvv}, while for $n>5$ we consider the $[2/3]$ Padé approximant of our $5$-loop result given by Eq. \eqref{fiveloops}. The latter was selected as it provides the closest agreement with the available bootstrap results for $n \le 6$. Note that for $n=6$ the NLO calculation fails as the $\ell=2$ stability angle is complex.

\begin{figure}[t!]
\centering
\includegraphics[width=0.42\textwidth]{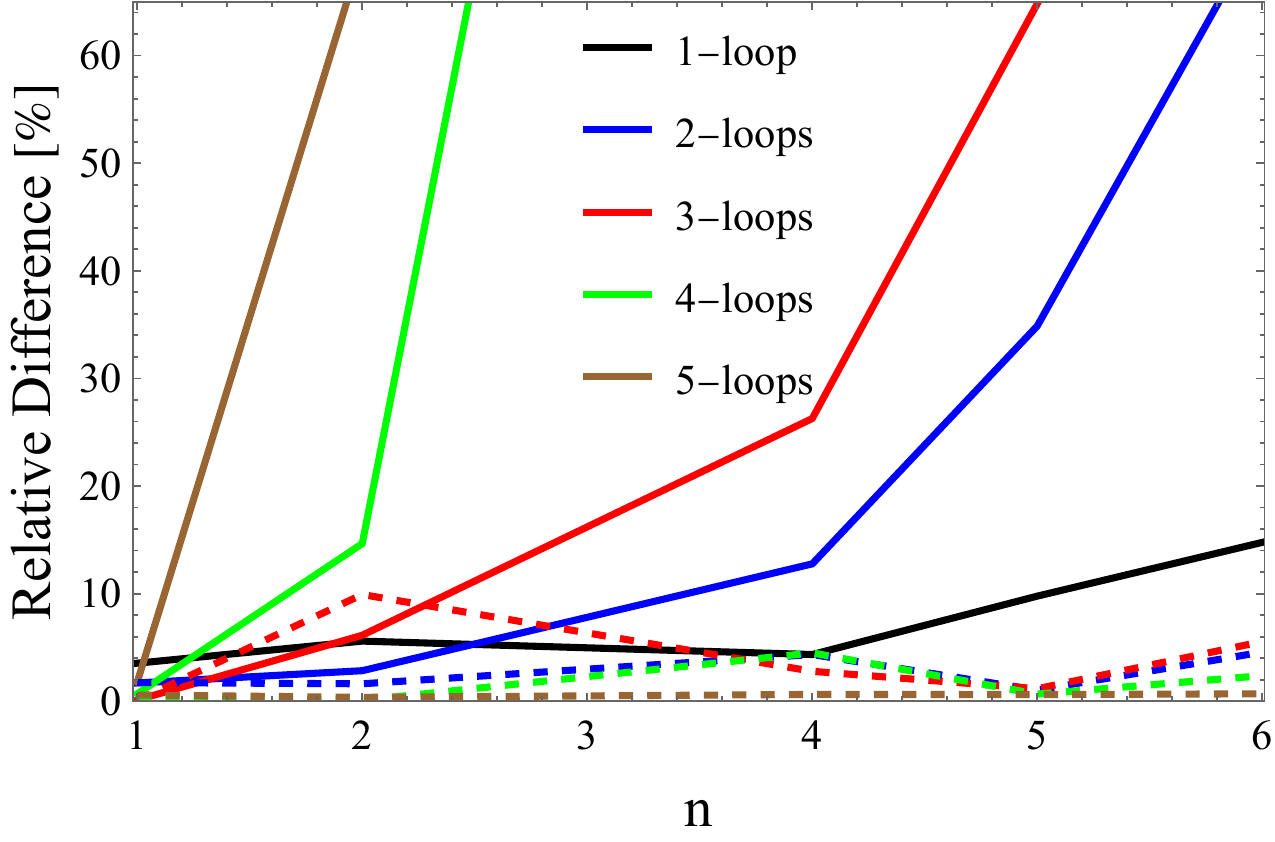}
	\caption{\emph{Solid lines}: Relative difference for $\Delta_n$ (in $\%$) between the bootstrap \cite{Simmons-Duffin:2016wlq, Henriksson:2022gpa} and the $\epsilon$-expansion truncated to various loop orders. \emph{Dashed lines}:  same quantity evaluated for the corresponding Padé approximants.} 
	\label{epserror}
\end{figure}

As it should be evident from the figure, the resummed $\epsilon$-expansion is in agreement with the available bootstrap results, and so no breakdown of the resummed perturbative theory is observed for $n<6$. To understand this observation, in Fig. \ref{epserror} we compare the results of the $\epsilon$-expansion truncated to various loop orders with and without Padé resummation. Clearly, resummation methods play a key role in improving the convergence of the $\epsilon$-expansion for larger $n$. In fact, the results of the $\epsilon$-expansion without resummation indicate that the perturbative series is an asymptotic one controlled by the product $\epsilon n$. In particular, for $n>2$, the optimal truncation order, and so the best possible result, is achieved at the $1$-loop level. Conversely, the sequence of Padé approximants exhibits convergence up to the $5$-loop level for any $n\le 6$.
From Fig.~\ref{Moneyplot}, it is clear that the semiclassical expansion performs well even when extrapolated to both $\epsilon = 1$ and low values of $n$, where numerical results are available. Crucially, the agreement with bootstrap results strengthens as $n$ grows
( from $\sim14\%$ for \(n=1\) to $\sim6\%$ for $n=5$). Moreover, we observe that for $n<12$ both the convergence of the semiclassical expansion and the agreement with the $\epsilon$-expansion, which itself is known to provide reliable estimates in this model, improve systematically as $n$ increases. This no longer holds for $n>12$, where the LO semiclassical results are closer to the $\epsilon$-expansion than the NLO ones, so the comparison ceases to be meaningful. Of course, it would be interesting to obtain precise numerical estimates for $n>6$ to test the actual accuracy of the semiclassical predictions in $d=3$. To this end, we collected our results for $n\le 16$ in App.~\ref{app2}.

\vskip .2cm
We conclude by remarking that while we focused on the Ising CFT, our framework can be employed to solve for the dynamics of composite operators stemming from different QFTs with applications ranging from condensed matter physics to particle physics and cosmology. For instance, we expect that gauge and Yukawa interactions can be incorporated within the same framework, allowing for the determination of the anomalous dimensions of composite operators that reduce, in the free field theory limit, to Higgs composites of the schematic form $\partial^s \Box^p(H^\dagger H)^n$.
In an interacting theory, such operators generally mix with operators built from fermionic and gauge degrees of freedom. Nevertheless, the semiclassical approach is expected to provide direct access to the eigenvalues of the anomalous dimension matrix, which correspond to the physical energy levels on the cylinder. This would provide a natural extension of our previous work on the Standard Model \cite{Antipin:2023tar} beyond the charged sectors with immediate impact for RG functions of higher-dimensional operators. Further interesting generalizations include composite operators involving higher powers of gauge-field strengths as well as fermionic operators of the schematic form $\left(\bar \psi \psi \right)^n$. In the latter case, however, additional complications are expected due to the enlarged operator basis generated by the gamma-matrix algebra.

\section*{Acknowledgments} 
The authors would like to thank A. V. Bednyakov, D. Orlando, and S. Reffert for useful comments. The work of G.M. and F.S. is partially supported by the Carlsberg Foundation, semper ardens grant CF22-0922. The work of J.B. is supported by the Swiss National Science Foundation under grant number 200021\_219267. 
\appendix
\section{Semiclassical calculation} \label{app1}
\begin{figure}[H]
\centering
\includegraphics[width=0.49\textwidth]{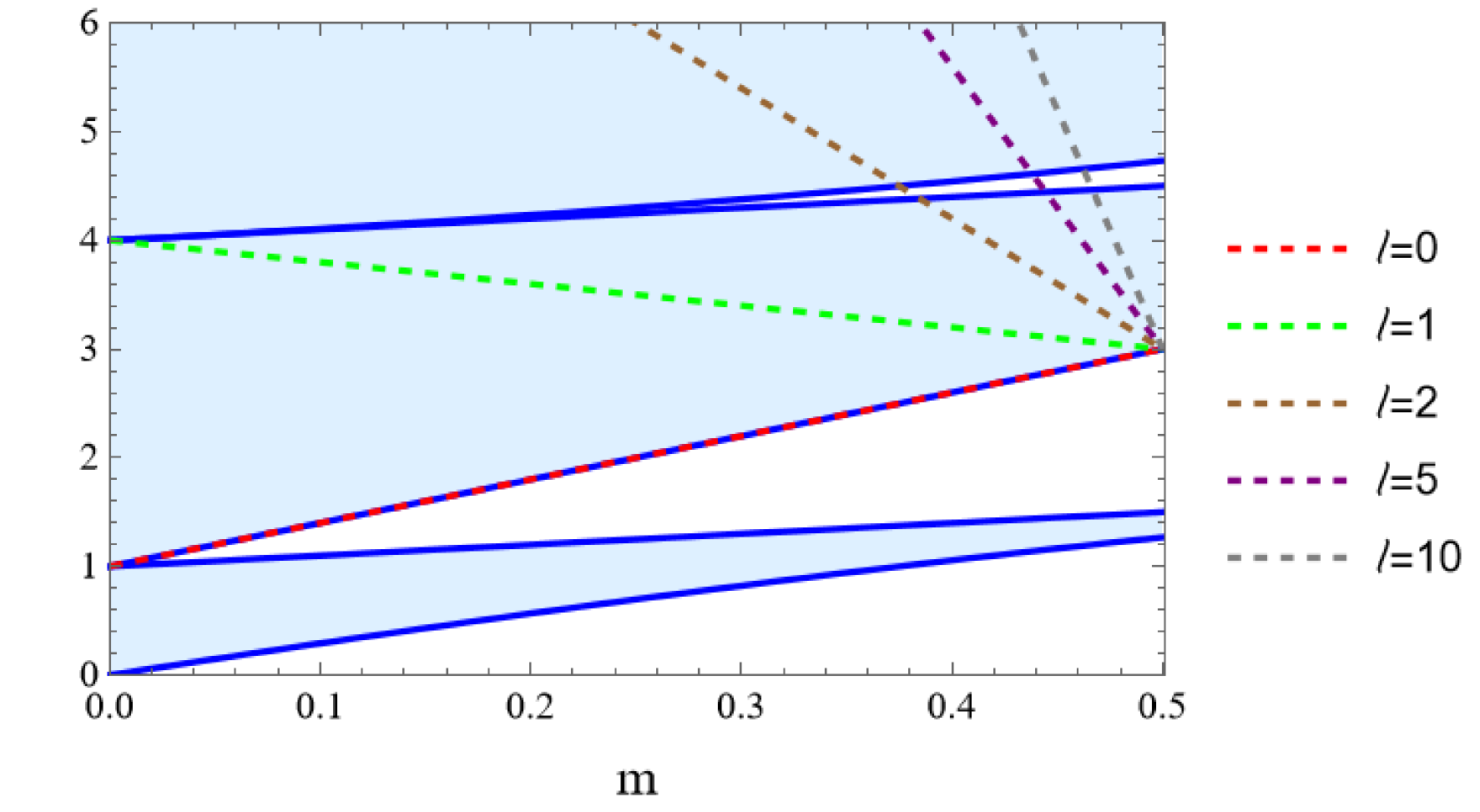} 
	\caption{Comparison of the band structure of the Lam\'e operator $L_2$ (allowed bands colored in light blue) and $\Lambda_\ell$ for $\ell=0,1,2,5,10$ (dashed lines) as a function of $m$. Complex stability angles arise when $\Lambda_\ell$ takes values inside the forbidden zones i.e., when dashed lines are inside the uncolored zones. 
    } 
	\label{plotanni60}
\end{figure}

 We here provide a derivation of our main result in Eq.~\eqref{labello}. The Lagrangian on the cylinder reads
\begin{equation} \label{Lagcyl}
    \mathcal{L}^{(\text{cyl})} =\frac{1}{2}(\partial \phi)^2 -\frac{\mu ^2 }{2}  \phi^2-\frac{\lambda}{4}\phi^4 \ ,
\end{equation}
with $\mu = \left( \frac{d-2}{2r}\right)$ the conformal mass. Assuming a spatially homogeneous field configuration, the time-dependent equation of motion for $\braket{\phi} = v(t)$ takes the form of the quartic anharmonic oscillator
\begin{equation}
\frac{d^2v}{dt^2} +\mu^2 v + \lambda v^3 = 0 \ ,
\label{eom}
\end{equation}
whose solution is \cite{Sanchez}
\begin{equation} \label{vevo}
v(t) = \, \mu \sqrt{\frac{2 m}{\lambda (1-2 m)}}\, \cn\left(\frac{\mu \ t}{\sqrt{1-2 m}}|m\right) \,, \quad 0<m<1/2
\end{equation}
where $\cn(\cdot|m) $ is the Jacobi elliptic cosine function. The period of the solution is 
\be
{\cal T} = \frac{4}{\mu} \sqrt{1-2m} \KK(m) \ , 
\ee
where $\KK(m)$ denotes the complete elliptic integral of the first kind. The associated energy levels to the next-leading-order in the semiclassical expansion can be derived following \cite{Gutzwiller:1971fy,dhn,Dashen:1974ci,Beccaria:2010zn} 
\begin{align} 
E_{n,\{q_\ell\}} = E_{cl}   -\frac{\lambda_*}{2} \beta _0  E_{cl}  +\frac{1}{\T}\sum_{\nu_\ell>0}\left(q_{\ell}+\frac{1}{2}\right) \nu_{\ell}\  .\label{pou}    
\end{align}
Here $E_{cl}$ is the classical energy of the solution \eqref{vevo} and $\beta_0 = \frac{9}{8\pi^2}$ is the $1$-loop coefficient of the $\beta$ function. The sum runs over all the positive stability angles $\nu_\ell$, with $\ell \in \mathbb{N}$, which emerge when analyzing the stability of the classical orbits by considering fluctuations around them. Mathematically, $e^{\pm i \nu_\ell}$ are the eigenvalues of the monodromy matrix associated with the eigenvalue equation of the fluctuation operator  \cite{floquet, Gutzwiller:1971fy}. The detailed derivation of Eq.~\eqref{pou} is lengthy and will be provided in a companion work \cite{Antipin:2025rsr}. The state-operator correspondence leads to 
\be \label{finafinal1}
  n C_0/r = E_{cl}\,, \quad  C_1/r =  -\frac{\lambda_*}{2} \beta _0  E_{cl} +\frac{1}{\T}\sum_{\nu_\ell>0}\left(q_{\ell}+\frac{1}{2}\right) \nu_{\ell}  \ .
\ee
$ E_{cl}$ and $\nu_\ell$ are functions of $m$ where the latter is related to $\lambda n$ by inverting the Bohr-Sommerfeld quantization condition $I = 2\pi n$ where the action variable $I$ reads
\begin{align}
I &= \oint \Pi d \phi =\Omega_{d-1} r^{d-1} \int_{0}^{{\cal T}} \left(\frac{dv}{dt} \right)^2 \,dt && \\ &= \frac{16 \pi^2  }{3  \lambda (1-2 m)^{3/2}} \left[ (2m-1)\ELE(m) + (1-m)\KK(m) \right]\,.\nonumber
\end{align}
Here $\ELE(m)$ is the complete elliptic integral of the second kind, and we introduced the field momentum $\Pi =\Omega_{d-1} r^{d-1}\dot \phi$. The classical energy yields $C_0$ \cite{Antipin:2024ekk}
\begin{equation}
  n C_{0} =r E_{\rm cl} =  \frac{2\pi^2  m \,(1-m) } {\lambda \,(1-2 m)^2} \ .
\end{equation}
To determine $C_1$, we expand the action around the classical trajectory as $\phi = v(t) + \eta(\vec x,t)$ obtaining the quadratic Lagrangian 
\be
\mathcal{L}_2 =  \frac{1}{2}\eta \cO_2 \eta \,, \quad \cO_2 = -\partial _t^2+\Delta _{S^{d-1}}-\mu ^2-   3 \lambda \, v^2(t)  \,,
\ee
with $\Delta _{S^{d-1}}$ the Laplacian on $S^{d-1}$. By changing variable as $z= \frac{\mu t}{\sqrt{1-2 m}}$ we  rewrite the fluctuation operator as 
\be \label{rescaled}
\cO_{2} = \frac{\mu^2 }{1-2m} L_2 \,,\quad L_{2} = -\partial_z^2+6 \ m \ \sn^2(z|m)-\Lambda_\ell \,,
\ee
where $\sn(z|m)$ is the Jacobi elliptic sine function and
\be \label{lambda12}
\Lambda_\ell = 6 m +  (1-2 m)  \left(1 + \frac{2 \ell}{d-2}\right)^2 \,,
\ee
includes $\Delta_{S^{d-1}}$ in terms of its  eigenvalues $J_{\ell}^2=\ell (\ell+d-2)/r^2$. Remarkably, $L_2$ takes the form of the $2$-gap Lam\'e operator whose stability angles are given by \cite{book1, Pawellek:2008st}
\be 
 \nu_{\ell} = 2\pi - 4 i \KK(m)  ( Z(\alpha_+\,|\,m)+\, Z(\alpha_-\,|\,m)) \ ,
\ee
where $Z$ denotes the Jacobi Zeta function and $\alpha_\pm$ solve 
\begin{align}
 \sn^2(\alpha_\pm |\,  m)&=  \pm\frac{1}{2m}\sqrt{\frac{4}{3}(1-m+m^2)-\frac{1}{3}\left(\Lambda_\ell-2(1+m)\right)^2} \nonumber \\ & + \frac{4(1+m)-\Lambda_\ell}{6m} \,.
\end{align}
When the spectral parameter $\Lambda_\ell$ takes value inside one of the two gap region of $L_2$, which are given by 
\be
\left\{ 1+ m, 1+4m \right\}\,, \quad \left\{ 4+ m , 2 \left(m+\sqrt{1-m (1-m)}+1\right)\right\}\,, \ee 
the stability angles develop a nontrivial imaginary part. By solving this condition for $\ell$ one obtains Eq. \eqref{ranger}. The known \cite{arscott} band structure of $L_2$ as a function of $0<m<1/2$ is shown in Fig. \ref{plotanni60} along with the values of $\Lambda_\ell$ for various values of $\ell$. 
\vskip 1em
Finally, by regularizing the sum over $\ell$ in Eq. \eqref{finafinal1}, renormalizing the result via dimensional regularization, and setting $d=4$ at the end of the calculation, we arrive at the expression for $C_1$ presented in the main text. 

\section{Semiclassical results in three dimensions}  \label{app2}
In \autoref{tabella}, we compare our semiclassical predictions for $\Delta_n$ to the results obtained via the conformal bootstrap and the $\epsilon$-expansion for $n=1,2, \dots, 16$.
\begin{table}[H]
\centering
\begin{tabular}{lcccc}
\toprule
$n$  & \textbf{Bootstrap}  & \textbf{$\epsilon$-expansion} &\textbf{LO} &\textbf{NLO}\\
\midrule
$1$  & $0.5181489(10)$  \cite{Simmons-Duffin:2016wlq}   & $0.5181(6)$ \cite{Kompaniets:2017yct} & $1.133$  & $0.5960$ \\
$2$  & $1.412625(23)$  \cite{Simmons-Duffin:2016wlq}   & $1.4121(6)$  \cite{Schnetz:2016fhy,Shalaby:2020xvv} & $2.459$  & $1.552$ \\
$4$  & $3.82968(23)$  \cite{Simmons-Duffin:2016wlq}   & $3.8231(5) \cite{Schnetz:2016fhy,Shalaby:2020xvv} $ & $5.509$ & $4.126$  \\
$5$ & $5.2906(11)$  \cite{Simmons-Duffin:2016wlq}  & $5.257$ & $7.190$ & $5.660$ \\
$6$ & $6.8956(43)$  \cite{Simmons-Duffin:2016wlq}   & $6.862$ & $8.956$ & X \\
$7$  & X  & $8.627$ & $10.80$ & $9.268$ \\
$8$  & X & $10.56$ & $12.72$ & $11.17$ \\
$9$  &X & $12.67$ & $14.70$  & $13.22$ \\
$10$& X & $15.0$ & $16.74$  & $15.38$ \\ 
$11$&X & $17.51$ & $18.85$ & $17.61$\\ 
$12$&X & $20.27$ & $21.00$ & $19.95$  \\ 
$13$&X & $23.31$ & $23.21$  & $22.37$\\ 
$14$&X & $26.67$ & $25.47$ & $24.87$ \\ 
$15$&X & $30.38$ & $27.77$ & $27.44$ \\ 
$16$ & X & $34.51$ & $30.12$ & $30.08$ \\ 
\bottomrule
\end{tabular}
\caption{$\Delta_n$ for $n \le 16$ obtained via numerical bootstrap,  $\epsilon$-expansion,  and semiclassics at LO and NLO.}
\label{tabella}
\end{table}


\vspace{1.5em}


\begin{thebibliography}{1}

\bibitem{Hellerman:2015nra}
S.~Hellerman, D.~Orlando, S.~Reffert and M.~Watanabe,
``On the CFT Operator Spectrum at Large Global Charge,''
JHEP \textbf{12} (2015), 071
doi:10.1007/JHEP12(2015)071
[arXiv:1505.01537 [hep-th]].

\bibitem{Komargodski:2012ek}
Z.~Komargodski and A.~Zhiboedov,
``Convexity and Liberation at Large Spin,''
JHEP \textbf{11} (2013), 140
doi:10.1007/JHEP11(2013)140
[arXiv:1212.4103 [hep-th]].

\bibitem{Rattazzi:2008pe}
R.~Rattazzi, V.~S.~Rychkov, E.~Tonni and A.~Vichi,
``Bounding scalar operator dimensions in 4D CFT,''
JHEP \textbf{12} (2008), 031
doi:10.1088/1126-6708/2008/12/031
[arXiv:0807.0004 [hep-th]].


\bibitem{Badel:2019oxl}
G.~Badel, G.~Cuomo, A.~Monin and R.~Rattazzi,
``The Epsilon Expansion Meets Semiclassics,''
JHEP \textbf{11} (2019), 110
doi:10.1007/JHEP11(2019)110
[arXiv:1909.01269 [hep-th]].

\bibitem{Buchmuller:1985jz}
W.~Buchmuller and D.~Wyler,
``Effective Lagrangian Analysis of New Interactions and Flavor Conservation,''
Nucl. Phys. B \textbf{268} (1986), 621-653
doi:10.1016/0550-3213(86)90262-2

\bibitem{Grzadkowski:2010es}
B.~Grzadkowski, M.~Iskrzynski, M.~Misiak and J.~Rosiek,
``Dimension-Six Terms in the Standard Model Lagrangian,''
JHEP \textbf{10} (2010), 085
doi:10.1007/JHEP10(2010)085
[arXiv:1008.4884 [hep-ph]].

\bibitem{Brivio:2017vri}
I.~Brivio and M.~Trott,
``The Standard Model as an Effective Field Theory,''
Phys. Rept. \textbf{793} (2019), 1-98
doi:10.1016/j.physrep.2018.11.002
[arXiv:1706.08945 [hep-ph]].

\bibitem{Goldberg:1990qk}
H.~Goldberg,
``Breakdown of perturbation theory at tree level in theories with scalars,''
Phys. Lett. B \textbf{246} (1990), 445-450
doi:10.1016/0370-2693(90)90628-J

\bibitem{Cornwall:1990hh}
J.~M.~Cornwall,
``On the High-energy Behavior of Weakly Coupled Gauge Theories,''
Phys. Lett. B \textbf{243} (1990), 271-278
doi:10.1016/0370-2693(90)90850-6

\bibitem{Son:1995wz}
D.~T.~Son,
``Semiclassical approach for multiparticle production in scalar theories,''
Nucl. Phys. B \textbf{477} (1996), 378-406
doi:10.1016/0550-3213(96)00386-0
[arXiv:hep-ph/9505338 [hep-ph]].

\bibitem{Khoze:2014zha}
V.~V.~Khoze,
``Multiparticle Higgs and Vector Boson Amplitudes at Threshold,''
JHEP \textbf{07} (2014), 008
doi:10.1007/JHEP07(2014)008
[arXiv:1404.4876 [hep-ph]].

\bibitem{Antipin:2024ekk}
O.~Antipin, J.~Bersini and F.~Sannino,
``Exact results for scaling dimensions of neutral operators in scalar conformal field theories,''
Phys. Rev. D \textbf{111} (2025) no.4, L041701
doi:10.1103/PhysRevD.111.L041701
[arXiv:2408.01414 [hep-th]].

\bibitem{Kehrein:1992fn}
S.~Kehrein, F.~Wegner and Y.~Pismak,
``Conformal symmetry and the spectrum of anomalous dimensions in the N vector model in four epsilon dimensions,''
Nucl. Phys. B \textbf{402} (1993), 669-692
doi:10.1016/0550-3213(93)90124-8

\bibitem{Kehrein:1994ff}
S.~K.~Kehrein and F.~Wegner,
``The Structure of the spectrum of anomalous dimensions in the N vector model in (4-epsilon)-dimensions,''
Nucl. Phys. B \textbf{424} (1994), 521-546
doi:10.1016/0550-3213(94)90406-5
[arXiv:hep-th/9405123 [hep-th]].

\bibitem{Cuomo:2024fuy}
G.~Cuomo, L.~Rastelli and A.~Sharon,
``Moduli spaces in CFT: large charge operators,''
JHEP \textbf{09} (2024), 185
doi:10.1007/JHEP09(2024)185
[arXiv:2406.19441 [hep-th]].

\bibitem{Gutzwiller:1971fy}
M.~C.~Gutzwiller,
``Periodic orbits and classical quantization conditions,''
J. Math. Phys. \textbf{12} (1971), 343-358
doi:10.1063/1.1665596

\bibitem{dhn}
R.~F.~Dashen, B.~Hasslacher and A.~Neveu,
``The Particle Spectrum in Model Field Theories from Semiclassical Functional Integral Techniques,''
Phys. Rev. D \textbf{11} (1975), 3424
doi:10.1103/PhysRevD.11.3424

\bibitem{Dashen:1974ci}
R.~F.~Dashen, B.~Hasslacher and A.~Neveu,
``Nonperturbative Methods and Extended Hadron Models in Field Theory 1. Semiclassical Functional Methods,''
Phys. Rev. D \textbf{10} (1974), 4114
doi:10.1103/PhysRevD.10.4114

\bibitem{Beccaria:2010zn}
M.~Beccaria, G.~V.~Dunne, G.~Macorini, A.~Tirziu and A.~A.~Tseytlin,
``Exact computation of one-loop correction to energy of pulsating strings in $AdS_5 x S^5$,''
J. Phys. A \textbf{44} (2011), 015404
doi:10.1088/1751-8113/44/1/015404
[arXiv:1009.2318 [hep-th]].


\bibitem{Sanchez}
A. Martín Sánchez and J. Díaz Bejarano,
``Quantum anharmonic symmetrical oscillators using elliptic functions,''
1986 J. Phys. A: Math. Gen. 19 887
doi:10.1088/0305-4470/19/6/019

\bibitem{floquet}
 W.~Magnus and S.~Winkler,
``Hill’s Equation,''
(Wiley, New York, 1966)

\bibitem{Antipin:2025rsr}
O.~Antipin, J.~Bersini, J.~Hafjall, G.~Muco and F.~Sannino,
``Semiclassical canovaccio for composite operators,''
JHEP \textbf{05} (2026), 137
doi:10.1007/JHEP05(2026)137
[arXiv:2512.23539 [hep-th]].

\bibitem{book1}
E.~T.~Whittaker, G.~N.~Watson,
``A Course of Modern Analysis,''
 Cambridge University Press (1927)

\bibitem{Pawellek:2008st}
M.~Pawellek,
``Quantum mass correction for the twisted kink,''
J. Math. Phys. \textbf{42} (2009), 045404
doi:10.1088/1751-8113/42/4/045404
[arXiv:0802.0710 [hep-th]].

\bibitem{arscott}
F.~M.~Arscott,
``Periodic Differential Equations: An Introduction to Mathieu, Lamé, and Allied Functions,''
Pergamon Press (1964)

\bibitem{ancillary}
The ancillary Mathematica notebook Ising\_Data.nb contains explicit seven-loop results for the spectrum of scaling dimensions in the critical $\phi^4$ theory.

\bibitem{Henriksson:2025hwi}
J.~Henriksson, F.~Herzog, S.~R.~Kousvos and J.~Roosmale Nepveu,
``Multiloop spectra in general scalar EFTs and CFTs,''
Phys. Lett. B \textbf{874} (2026), 140235
doi:10.1016/j.physletb.2026.140235
[arXiv:2507.12518 [hep-ph]].

\bibitem{Schnetz:2016fhy}
O.~Schnetz,
``Numbers and Functions in Quantum Field Theory,''
Phys. Rev. D \textbf{97} (2018) no.8, 085018
doi:10.1103/PhysRevD.97.085018
[arXiv:1606.08598 [hep-th]].

\bibitem{Schnetz:2022nsc}
O.~Schnetz,
``{\ensuremath{\phi}}4 theory at seven loops,''
Phys. Rev. D \textbf{107} (2023) no.3, 036002
doi:10.1103/PhysRevD.107.036002
[arXiv:2212.03663 [hep-th]].

\bibitem{Bednyakov:2021ojn}
A.~Bednyakov and A.~Pikelner,
``Six-loop beta functions in general scalar theory,''
JHEP \textbf{04} (2021), 233
doi:10.1007/JHEP04(2021)233
[arXiv:2102.12832 [hep-ph]].

\bibitem{Kompaniets:2017yct}
M.~V.~Kompaniets and E.~Panzer,
``Minimally subtracted six loop renormalization of $O(n)$-symmetric $\phi^4$ theory and critical exponents,''
Phys. Rev. D \textbf{96} (2017) no.3, 036016
doi:10.1103/PhysRevD.96.036016
[arXiv:1705.06483 [hep-th]].

\bibitem{Henriksson:2022rnm}
J.~Henriksson,
``The critical O(N) CFT: Methods and conformal data,''
Phys. Rept. \textbf{1002} (2023), 1-72
doi:10.1016/j.physrep.2022.12.002
[arXiv:2201.09520 [hep-th]].

\bibitem{Dunne:2015eaa}
G.~V.~Dunne and M.~{\"U}nsal,
``What is QFT? Resurgent trans-series, Lefschetz thimbles, and new exact saddles,''
PoS \textbf{LATTICE2015} (2016), 010
doi:10.22323/1.251.0010
[arXiv:1511.05977 [hep-lat]].

\bibitem{Simmons-Duffin:2016wlq}
D.~Simmons-Duffin,
``The Lightcone Bootstrap and the Spectrum of the 3d Ising CFT,''
JHEP \textbf{03} (2017), 086
doi:10.1007/JHEP03(2017)086
[arXiv:1612.08471 [hep-th]].

\bibitem{Henriksson:2022gpa}
J.~Henriksson, S.~R.~Kousvos and M.~Reehorst,
``Spectrum continuity and level repulsion: the Ising CFT from infinitesimal to finite {\ensuremath{\varepsilon}},''
JHEP \textbf{02} (2023), 218
doi:10.1007/JHEP02(2023)218
[arXiv:2207.10118 [hep-th]].

\bibitem{Shalaby:2020xvv}
A.~M.~Shalaby,
``Critical exponents of the O(N)-symmetric $\phi ^4$ model from the $\varepsilon ^7$ hypergeometric-Meijer resummation,''
Eur. Phys. J. C \textbf{81} (2021) no.1, 87
[arXiv:2005.12714 [hep-th]].


\bibitem{Antipin:2023tar}
O.~Antipin, J.~Bersini, P.~Panopoulos, F.~Sannino and Z.~W.~Wang,
``Infinite order results for charged sectors of the Standard Model,''
JHEP \textbf{02} (2024), 168
doi:10.1007/JHEP02(2024)168
[arXiv:2312.12963 [hep-ph]].


\end{thebibliography}
\end{document}